# A Survey on Service Composition Middleware in Pervasive Environments

**Noha Ibrahim[1], Frédéric Le Mouël[2]**

**[1] Grenoble Informatics Laboratory
Grenoble, France
noha.ibrahim@imag.fr**

**[2] Université de Lyon, INRIA, INSA-Lyon, CITI
Lyon, France
frederic.le-mouel@insa-lyon.fr**

## Abstract

The development of pervasive computing has put the light on a challenging problem: how to dynamically compose services in heterogeneous and highly changing environments? We propose a survey that defines the service composition as a sequence of four steps: the translation, the generation, the evaluation, and finally the execution. With this powerful and simple model we describe the major service composition middleware. Then, a classification of these service composition middleware according to pervasive requirements - interoperability, discoverability, adaptability, context awareness, QoS management, security, spontaneous management, and autonomous management - is given. The classification highlights what has been done and what remains to do to develop the service composition in pervasive environments.

## 1. Introduction

Middleware are enabling technologies for the development, execution and interaction of applications. These software layers are standing between the operating systems and applications. They have evolved from simple beginnings - hiding network details from applications - into sophisticated systems that handle many important functionalities for distributed applications - providing support for distribution, heterogeneity and mobility. SOA middleware[2] is a programming paradigm that uses ``services'' as the unit of computer work. Service-oriented computing enables the development of loosely coupled systems that are able to communicate, compose and evolve in an open, dynamic and heterogeneous environment. A service-oriented system comprises software systems that interact with each other through well-defined interfaces.

If middleware were designed to help manage the complexity and heterogeneity inherent in distributed systems, one can imagine the new role middleware has to play in order to respect the evolution from distributed and mobile computing to pervasive one. Hardly a day passes without some new evidence of the proliferation of portable computers in the marketplace, or of the growing demand for wireless communication. Support for mobility has been the focus of number of experimental systems, researches and commercial products, and that since several decades. The mission of mobile computing is to allow users to access any information using any device over any network at any time. When this access becomes to every information using every device over every network at every time, one can say that mobile computing has evolved to what we now call pervasive computing[13].

In pervasive environments where SOA has been adopted, functionalities are more and more modeled as services, and published as interfaces. The proliferation of new services encourages the applications to use these latter, all combined together. In this case, we speak of a composite service. The process of developing a composite service is called service composition[7]. Composing services together is the new challenge awaiting the SOA middleware[2] meeting the pervasive environments[13]. Indeed, the variety of service providers in a pervasive environment, and the heterogeneity of the services they provide require applications and users of these kind of environments to develop models, techniques and algorithms in order to compose services and execute them. The service composition needs to follow some

requirements[19][33][34] in order to resolve the challenges brought by pervasivity.

Several surveys[5][7][22][31][33] dealt with service composition. Many of them[7][31] classified the middleware under exclusive criteria such as manual versus automated, static versus dynamic, and so on. Others[5][22][33] classified the service composition middleware under different domains such as artificial intelligence, formal methods, and so on. But none of these surveys proposed a generic reference model to describe the service composition middleware in pervasive environments.

In this article, we propose:

- a generic service composition middleware model, the SCM model, a novel way to describe the service composition problem in pervasive environments,
- a description of six middleware architectures using the SCM model as a backbone and highlighting the strength and weakness of each middleware,
- and finally, a classification of these latter under pervasive requirements identified by the literature to be essential for service composition in pervasive environments.

The outlines are as follows. In section 2, we define the service composition middleware (SCM) model and explain its modules. In section 3, we describe six service composition middleware by mapping their architecture to the SCM model. In section 4, we classify these middleware according to the pervasive requirements we identify. Finally, section 5 concludes our work and gives research directions to the service composition problem.

## 2. SCM: Service Composition Middleware Model

Based on several studies[22][24] that resolve the service composition process problem into several fundamental problems, we define a service composition middleware as a framework providing tools and techniques for composing services. We define a service composition middleware model, SCM model, as an abstract layer, general enough to describe all existing service composition middleware. The SCM model is at a high-level of abstraction, without considering a particular service technology, language, platform or algorithm used in the composition process. The aim of this definition is to give the basis to discuss similarities and differences, advantages and disadvantages of all available service composition middleware and to highlight the nowadays existing lacks concerning the service composition problem in pervasive environments.

As depicted Figure 1, the SCM interacts with the application layer by receiving functionality requests from users or applications[5][7]. SCM needs to respond to the functionality requests by providing services that fulfill the demand. These services can be atomic or composite. The Service Repository represents all the distributed service repository where services are registered. The SCM interacts with the Service Repository to choose services to compose.

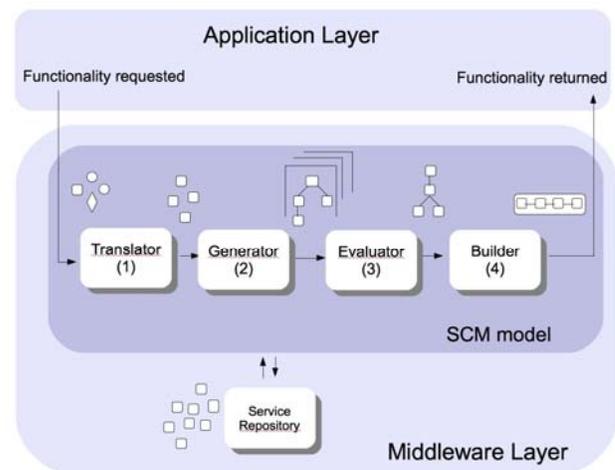

Figure 1 SCM model

The SCM is split into four components: the *Translator*, the *Generator*, the *Evaluator*, and the *Builder*. The process of service composition includes the following phases:

1. Applications specify their needed functionalities by sending requests to the middleware. These requests can be described with diverse languages or techniques. The request descriptions are translated to a system comprehensible language in order to be used by the middleware. Most systems distinguish between external specification languages and internal ones. The external ones are used to enhance the accessibility with the outside world, commonly the users. Users can hence express what they need or want in a relatively easy way, usually using semantics and ontologies. Internal specification corresponds more to a formal way of expressing things and uses specific languages, models, and logics, usually for SOA a generic service model. Some research[30] provide a translation mechanism of the available service technologies and service descriptions into one model. Others, such as SELF-SERV[25], propose a wrapper to provide a uniform access interface to services[8]. These middleware usually realize transformation from one model to another or from one technology to another. The technologies are predefined in advance and usually consist of the legacy ones. If new technology models appear in the environment, the *Translator* will need to be expanded to take these technologies into consideration. Another family of research[6][26] do not provide the *Translator* module as they use common model to describe all the services of the environment. They use common description languages such as DAML-S - recently called OWL-S[36], - for describing atomic services, composed services and user queries.

2.  Once translated, the request specification is sent to the *Generator*. The *Generator* will try to provide the needed functionalities by composing the available service technologies, and hence composing their functionalities. It tries to generate one or several composition plans with the same or different technology services available in the environment. It is quite common to have several ways to do a same requirement, as the number of available functionalities in pervasive environments is in expansion. Composing service is technically performed by chaining interfaces using a syntactically or semantically method matching. The interface chaining is usually represented as a graph or described with a specific language. Graph based approaches[8][10], represent the semantic matching between the inputs and outputs of service operations. It is a powerful technique as many algorithms can be applied upon graphs and hence optimize the service composition. Number of languages have been proposed in the literature to describe data structure in general and functionalities offered by devices in particular. If some languages are widely used, such as XML, and generic for multiple uses others are more specific to certain tasks as service composition, orchestration or choreography such as Business Process Execution Language (BPEL4WS or BPEL[4]) and OWL-S[36].

3.  The *Evaluator* chooses the most suitable composition plan for a given context. This selection is done from all the plans provided by the *Generator*. In pervasive environments, this evaluation depends strongly on many criteria like the application context, the service technology model, the quality of the network, the non functional service QoS properties, and so on. The evaluation needs to be dynamic and adaptable as changes may occur unpredictably and at any time. Two main approaches are commonly used: the rule-based planning[27][28][29] and the formal methods approach[6][10][12][30]. The rules evaluate whether a given composition plan is appropriate or not in the actual context. If rules were commonly used as an evaluation approach, their use lacks of dynamism proper to pervasive environments. A major problem of the evaluation approach is namely the lack of dynamic tools to verify the correctness - functional and non functional aspects - of the service composition plan. This aspect is at the main advantage of what most formal methods offer. The nowadays most popular and advanced technique to evaluate a given composition plan is the evaluation by formal methods (like Petri nets and process algebras like the Pi-calculus). Petri nets are a framework to model concurrent systems. Their main attraction is the natural way of identifying basic aspects of concurrent systems, both mathematically and conceptually. Petri nets are very commonly merged with composition languages such as BPEL[4] and OWL-S[36]. On the other hand, Automata or labeled transition systems are a well-known model underlying formal system specifications and are more and more used in the service composition process[30].

4.  The *Builder* executes the selected composition plan and produces an implementation corresponding to the required composite service. It can apply a range of techniques to realize the effective service composition. These techniques depend strongly on the service technology model we are composing and on the context we are evolving in. Once the composite service available, it can be executed by the application that required its functionality. In the literature, we distinguish different kinds of builders provided by the service composition middleware. Some builders are very basic and use only simple invocation in sequence to a list of services[17]. These services need to be

**IJCSI**

available otherwise the composition result is not certain. Others[35] provide complex discovery protocols adapted to the heterogeneous nature of the pervasive environments. The discovery takes in charge to find and choose the services taking part into the composition process and to choose contextually the most suitable ones if many are available. Finally some systems propose solutions not only located in the middleware layer but also in the networking one.

We argue that the SCM model is generic enough to describe the service composition process in pervasive environments. In the next section, we use the SCM model as a backbone for describing various middleware that do the service composition.

## 3. Service Composition Middleware in Pervasive Environments

In this section, we describe six middleware for service composition adapted for pervasive environments by mapping them to our SCM model. The chosen middleware are architectures, platforms or algorithms that propose solutions to the service composition problem: MySIM[17], PERSE[30], SeSCo[10], Broker[6], SeGSeC[8] and WebDG[12].

For each middleware, we describe the service composition runtime process, the prototypes developed and identify the four modules of our SCM model in their provided architectures.

### 3.1 MySIM: Spontaneous Service Integration for Pervasive Environment

MySIM[17] is a spontaneous middleware that integrate services in a transparent way without disturbing users and applications of the environment. Service integration is defined as being a service transformation from one service technology to another (*Translator*), a service composition and a service adaptation. MySIM selects services that are composable, generates composition plans (*Generator*), evaluate their QoS degrees (*Evaluator*) and implements new composite services in the environment (*Builder*). These new services publish well known interfaces but new implementations and better QoS. MySIM also proposes to adapt the application execution to the services available by redirecting the application call to services with better QoS.

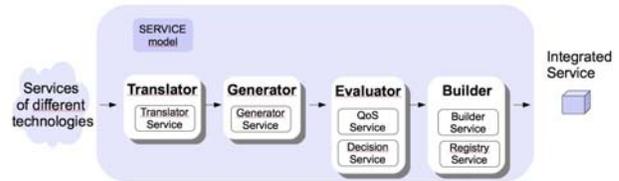

Figure 2 MySIM mapped to SCM

MySIM architecture is depicted under the SCM model in Figure 2. The *Translator* service transforms services into a generic Service model. The *Generator* service is responsible of the syntactic and semantic matching of the service operations for composition and adaptation issues. The *QoS* service evaluates the composition or substitution matching via non functional properties and the *Decision* service decides which services to compose or to substitute. Finally the *Builder* service implements the composite service, and the *Registry* service publishes its interfaces.

MySIM is implemented under the OSGi/Felix platform. It uses the reflexive techniques to do the syntactic interface matching and ontology online reasoner for the semantic matching. The service composition is technically done by generating new bundles (unit of deployment) that composes the services together. The results show the heavy cost of the semantic matching. The solution is interesting but solutions need to be found to make the spontaneous service integration scalable to large environments.

### 3.2 PERSE: Pervasive Semantic-aware Middleware

PERSE[30] proposes a semantic middleware, that deals with well known functionalities such as service discovery, registration and composition. This middleware provides a service model to support interoperability between heterogeneous both semantic and syntactic service description languages (*Translator*). The model further supports the formal specification of service conversations as finite state automata, which enables the automated reasoning about service behavior independently from the underlying conversation specification language. Hence, pervasive service conversations described with different service conversation languages (*Generator*) can be integrated (*Builder*) toward the realization of a user task. The model also supports the specification of service non-functional properties based on existing QoS models to meet the specific requirements of each pervasive application through the QoS aware Composition service (*Evaluator*).

**IJCSI**

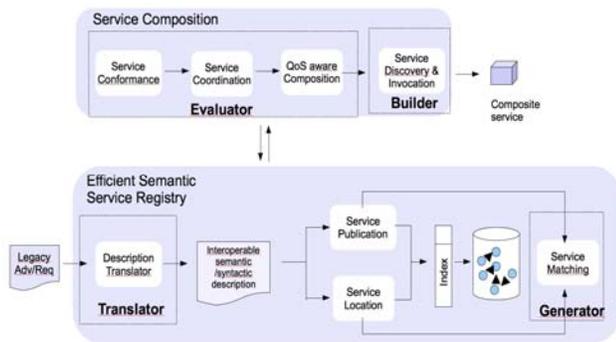

Figure 3 PERSE mapped to SCM

PERSE architecture is depicted under the SCM model in Figure 3. The *Evaluator* module is the most developed as it verifies the correctness of the composition plan and analyzes the service QoS before composing services. A *Translator* is also available to translate the legacy services into a common model semantically and syntactically described. The *Generator* semantically matches services. The *Builder* discovers the services in the environment and simply invoke them in sequence.

[30] have implemented a prototype of PERSE using Java 1.5. Selected legacy plugins have been developed for SLP using jSLP, UPnP[35] using Cyberlink, and UDDI using jUDDI. The efficiency of PERSE has been tested and proved in the cost evaluation of semantic service matching, the time to organize the semantic service registry, the time to publish and locate a semantic service description as well as the comparison of the scalability of the registry compared with a WSDL service registry, and finally the processing time for service composition with and without the support of QoS.

### 3.3 SeSCo: Seamless Service Composition

SeSCo[10] presents a service composition mechanism for pervasive computing. It employs the service-oriented middleware platform called Pervasive Information Communities Organization (PICO) to model and represent resources as services. The proposed service composition mechanism models services as directed attributed graphs, maintains a repository of service graphs, and dynamically combines multiple basic services into complex services (*Builder*). The proposed service composition mechanism constructs possible service compositions based on their semantic and syntactic descriptions (*Generator*). SeSCo[10] proposes a hierarchical service overlay mechanism based on a LATCH protocol (*Evaluator*). The hierarchical scheme of aggregation exploits the presence of heterogeneity through service cooperation. Devices with higher resources assist those with restricted resources in accomplishing service-related tasks such as discovery, composition, and execution.

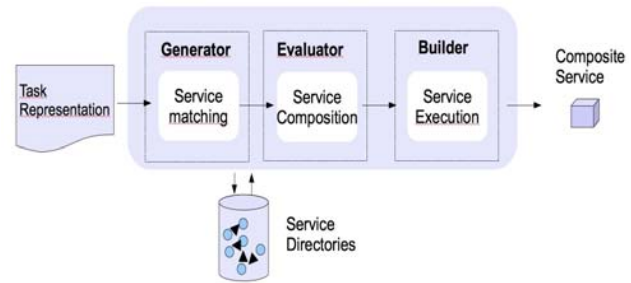

Figure 4 SeSCo mapped to SCM

SeSCo architecture is depicted under SCM model in Figure 4. No *Translator* module is provided and SeSCo uses the same language to present the user task and the composite service. The service matching is done on a semantic interface matching and the evaluation is upon the input/output matching correctness.

SeSCo[10] evaluated its approach by calculating the composition success ratio for different lengths of composition which is essentially the number of services that can be used to compose a single service. This evaluation shows the effect of limiting the length of the composition to a predefined number. If the service density is higher, even with a lower value of composition length, a successful composition can be achieved. However, at lower service densities, it might be necessary to allow higher composition lengths for better composition.

### 3.4 Broker Approach for Service Composition

Broker[6] presents a distributed architecture and associated protocols for service composition in mobile environments that take into consideration mobility, dynamic changing service topology, and device resources. The composition protocols are based on distributed brokerage mechanisms (*Evaluator*) and utilize a distributed service discovery process over ad-hoc network connectivity. The proposed architecture is based on a composition manager, a device that manages the discovery, integration (*Generator*), and execution of a composite request (*Builder*). Two broker selection-based protocols - dynamic one and distributed one - are proposed in order to distribute the composition requests to the composition managers available in the environment. These protocols depend on device-specific potential value, taking into account services available on the devices, computation and energy resources and the service topology of the surrounding vicinity.

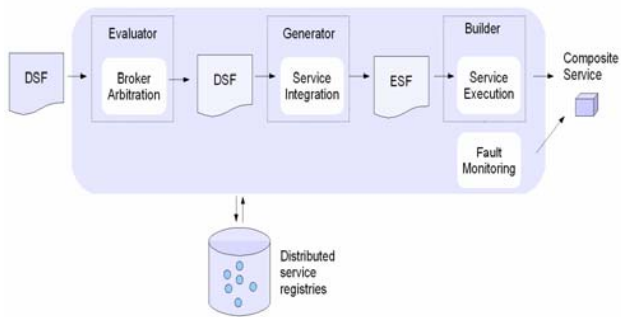

Figure 5 Broker mapped to SCM

Broker architecture is depicted under the SCM model in Figure 5. The *Broker arbitration* is the *Evaluator* module as it evaluates the available devices and decides to distribute the composition request, described in a special language (DSF), taking into account the device context. The evaluation is done here before the composition process. The *Service Integration* describes the composition sequence using a specific language (ESF) and pass it to the *Service Execution* (the *Builder*) to execute it.

Broker[6] implemented a protocol as part of a distributed event-based mobile network simulator, to test the two proposed broker arbitration protocols and the composition efficiency. Simulation results show that their protocols - especially the distributed approach - exceed the usual centralized broker composition in terms of composition efficiency and broker arbitration efficiency.

## 3.5 SeGSeC: Semantic Graph-Based Service Composition

SeGSeC[8] proposes an architecture that obtains the semantics of the requested service in an intuitive form (e.g. using a natural language) (*Tranlator*), and dynamically composes the requested service based on its semantics (*Generator*). To compose a service based on its semantics, the proposed architecture supports semantic representation of services - through a component model named Component Service Model with Semantics (CoSMoS) - discovers services required for composition - through a middleware named Component Runtime Environment (CoRE) - and composes the requested service based on its semantics and the semantics of the discovered services - through a service composition mechanism named Semantic Graph-Based Service Composition (SeGSeC).

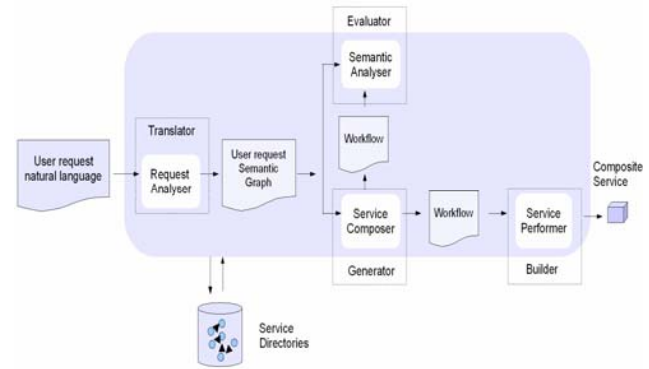

Figure 6 SeGSeC mapped to SCM

SeGSeC architecture is depicted under SCM model in Figure 6. The *Request Analyser* translates the user request into an internal system language using graph-based approach. The *Semantic Analyser* and *Service composer* produce the composition workflow ready to be executed by the *Service performer*. The workflow respects the semantic matching composition rules and the correctness is guaranteed via the *Evaluator* module.

SeGSeC[8] was evaluated according to the number of services deployed and the time needed to discover, match and compose services. Another set of evaluations took not only the number of deployed services but especially the number of operations these services implement. Their results show that SeGSeC performs efficiently when only a small number of services are deployed and that it scales to the number of services deployed if the discovery phase is done efficiently.

## 3.6 WebDG: Semantic Web Services Composition

WebDG[12] proposes an ontology-based framework for the automatic composition of web services. [12] presents an algorithm to generate composite services from high level declarative descriptions. The algorithm uses composability rules, in order to compare the syntactic and semantic features of web services to determine whether two services are composable.

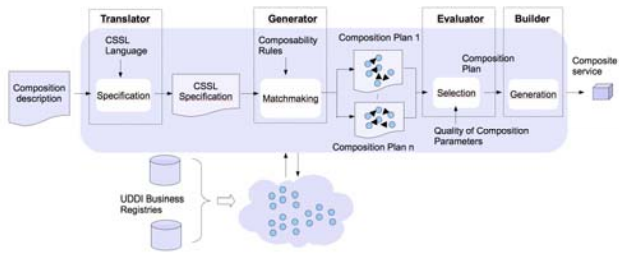

Figure 7 WebDG mapped to SCM

WebDG architecture is depicted under SCM model in Figure 7. The service composition approach is depicted under four phases of request specification (*Translator*), service description matchmaking (*Generator*), composition plan selection (*Evaluator*) and composite service generation (*Builder*).

A prototype implementation WebDG is provided and tested on a E-government Web service applications.

The WebDG evaluation aims to test the possibility of generating plans for a large number of service interfaces, the effectiveness and speed of the matchmaking algorithm, and the role of the selection phase (QoC parameters) in reducing the number of generated plans. The result show that most of the time is spent on checking message composability. On the other hand, a relatively low value of composition completeness generates more plans, each plan containing a small number of composable operations. In contrast, a high value of this ratio generates a smaller number of plans, each plan having more composable operations.

## 4. Classification of the Pervasive Service Composition Middleware

As shown above, the SCM model is generic enough to provide generic functional modules that describe the existing service composition middleware. We choose to classify the middleware – MySIM[17], PERSE[30], SeSCo[10], Broker[6], SeGSeC[8] and WebDG[12] - according to pervasive environment requirements. We first list and explain these pervasive requirements for service composition middleware, then a classification of these middleware is given.

### 4.1 Pervasive Requirements

Pervasive computing brought new challenges to distributed and mobile computing. We identify the following eight fundamental requirements for service composition in pervasive environments: interoperability, discoverability, adaptability, context awareness, QoS management, security, spontaneous management and autonomous management.

Interoperability is the ability of two or more systems or components to exchange information and to use the information that has been exchanged. Ubiquitous computing environments, quoting Mark Weiser's definition, consist of various kinds of computational devices, networks and collaborating software and hardware entities. Due to the large number of heterogeneous and cooperating parties, interoperability is required at all levels of ubiquitous computing. Service composition middleware need to take advantage of all the functionalities available in the surroundings, and for that they need to be interoperable.

Discoverability is a major issue for ubiquity and composition as devices and services need to be located and accessed before being composed. One of the fundamental challenges of distributed and highly dynamic environments is how the applications can discover the surrounding entities and, conversely, how the applications can be discovered by the other entities in the system. In a pervasive system, the execution environment of applications can be logically considered as a single container including all applications, other components, and resources. Moreover, the idea in distributed environments is that the resources can be accessed without any knowledge of where the resources or the users are physically located.

Adaptability is the ability of a software entity to adapt to the changing environment. Changes in applications' and users' requirements or changes within the network, may require the presence of adaptation mechanisms within the middleware. Moreover, adaptation is necessary when a significant mismatch occurs between the supply and demand of a resource. As the application's execution environment changes due to the user's mobility, the vital resources need to be substituted by corresponding resources in the new environment in order to ensure continuous operation. The requirement for adaptation is present on many different layers of a computing system.

Context awareness is the ability of pervasive middleware to be aware in terms of devices coming and leaving, functionalities offered and retrieved, quality of service changing, etc. They need to be aware of all these changes, in order to offer the best functionalities to applications regardless the context around. When considering context-aware systems in general, some common functionalities that are present in almost every system, can be identified: context sensing and processing, context information representation, and the applications that utilize the context information. In general, the context information can be divided into low-and high-level context information. Low-level context information can be collected using sensors in the system. Low-level context information sources can be combined or processed further to higher level context information.

QoS management is essential in dynamic environments, where connectivity is very variable. A pervasive middleware for service composition need to take the non functional parameters of applications and devices into consideration in order to provide viable and flexible composition plans and composite services. QoS parameters concern not only the services but also the devices where the execution is taking place. The composition execution need to rely on this parameter in order to take place in the best conditions. Not only the QoS of different services need to be compatible, but also the devices performing the composition need to respect certain characteristics and constraints.

Security mechanisms, such as authentication, authorization, and accounting (AAA) functions may be an important part of the middleware in order to intelligently control access to computer and network resources, enforcing policies, auditing network/user usage, etc. Another important aspect concerns privacy and trust in pervasive environments. In presence of unknown devices, middleware need to respect privacy of users, and provide trust mechanisms adapted to the ever changing nature of the environment.

Spontaneous management concerns the ability of a pervasive middleware to compose services independently of user and application requests. The middleware spontaneously composes services that are compatible together and produces a new composite service into the environment. The new service is registered and can publish its interfaces in order to be discovered and executed by applications. Spontaneous service composition is an interesting feature in pervasive environments, as services meet when the user encounter, and interesting composite service can be generated from these meetings, even though not required at that moment by users.

Autonomous Management concerns the ability for a pervasive middleware to control and manage its resources, functions, security and performance, in face of failures and changes, with little or no human intervention. The complexity of future ubiquitous computing environments will be such that it will be impossible for human administrators to perform their traditional functions of configuration management, performability management, and security management. Instead, one must resort to automate most of these management functions, allowing humans to concentrate on the definition and supervision of high-level management policies, while the middleware itself takes care of the translation of these high-level policies into automated control structures. The challenge is therefore to move from classical middleware support for configuration, performability and security management to support for self-configuration, self-tuning, self-healing and self-protecting capabilities.

We classify the service composition middleware – MySIM[17], PERSE[30], SeSCo[10], Broker[6], SeGSeC[8], and WebDG[12] - under the above requirements. For each middleware, we analyze its four modules - *Translator*, *Generator*, *Evaluator*, and *Builder* - and detail if they respect the pervasive requirements. The first section depicts the requirements that are fulfilled, at a certain extend, by the service composition middleware. The second section explains the requirements that are until now left behind. Our classification is given in Figure 8.

## 4.2 Service Composition Middleware Meeting Pervasive Requirements

In this section, we are interested in the pervasive requirements that are fulfilled by service composition middleware: discoverability, adaptability, context awareness, and QoS management.

If some pervasive requirements are relatively well fulfilled by the current composition middleware, others are still at a preliminary stage.

All middleware provide the discoverability and context awareness requirements. These requirements are intrinsic to every composition middleware wanting to evolve in dynamic and ever changing environment such as the pervasive environments. These requirements are essential when constructing and evaluating composition plans, but also when discovering and invoking services. Indeed, generating and evaluating composition plans must be contextual, as services can come and go at any time, and a given composition plan constructed at a certain time, need to be evaluated before execution, in case some changes have affected it. Hence, the context awareness is not only provided by the *Builder* but also by the *Generator* and *Evaluator* modules.

| Middleware | Requirements | Interoperability | Discoverability | Adaptability | Context awareness | QoS management | Spontaneous management |
|---|---|---|---|---|---|---|---|
| MYSIM | Translator | x | | | x | | |
| | Generator | | x | x | x | x | x |
| | Evaluator | | x | x | x | x | x |
| | Builder | | x | x | x | | |
| PERSE | Translator | x | | | x | | |
| | Generator | | x | x | x | x | |
| | Evaluator | | x | x | x | x | |
| | Builder | | x | | x | | |
| SeGSeC | Translator | x | | | | | |
| | Generator | | x | | x | | |
| | Evaluator | | | | x | | |
| | Builder | | x | | x | | |
| SeSCo | Translator | | | | | | |
| | Generator | | x | x | x | | |
| | Evaluator | | | | x | | |
| | Builder | | x | | x | x | |
| Broker | Translator | | x | | x | | |
| | Generator | | x | | x | | |
| | Evaluator | | x | | x | x | |
| | Builder | | x | x | x | | |
| WebDG | Translator | | x | | x | | |
| | Generator | | x | | x | | |
| | Evaluator | | x | | x | x | |
| | Builder | | | | x | | |

Figure 8 Service composition Middleware Classification

The adaptability requirement is fulfilled by four of the six classified middleware (MySIM[17], PERSE[30], SeSCo[10], and Broker[6]) for different SCM modules. The environmental changes, that affect a pervasive environment, such as devices coming and leaving, services being unavailable, require from the middleware special mechanisms in order to re-evaluate and adapt their service composition to these changes. As we can see, some middleware propose adaptation mechanisms, but this requirement is far from being fulfilled by all service composition middleware in the environment. In nowadays researches, adaptation is more considered as a field of research[35] than a requirement to fulfill. Adapting services can be seen as a way of integrating services into their new environments.

The QoS management requirement is fulfilled by five of the six classified middleware (MySIM[17], PERSE[30], SeSCo[10], Broker[6] and WebDG[11]). The modules that usually respect the QoS properties are the *Generator*, *Evaluator* and the *Builder*. The *Evaluator* relies on the service QoS parameters in order to choose the most suitable plan from all possible composition plans. QoS is especially relevant for stateful services. A plan composition of stateful services need to take QoS into account, as the resulting composition may not execute in case of severe incompatibilities in QoS between combined services. The *Builder* can analysis the QoS parameter in order to choose the devices and platforms where to execute the service composition, depending on power or memory properties, but also to choose services to compose depending on the devices they execute on. This requirement is especially considered in the development of multimedia applications in variable environments such as pervasive environments[16]. Indeed, composing services within multimedia applications imposes a rigorous respect of the QoS properties otherwise the whole application may not execute.

## 4.3 Service Composition Middleware Missing Pervasive Requirements

Nowadays service composition middleware present real lack in providing interoperability, spontaneous management, security mechanisms and autonomous management to service composition in pervasive environments.

The interoperablity requirement is more than left behind in nowadays service composition middleware. Figure 8 shows that only three middleware (MySIM[17], PERSE[30] and SeGSeC[8]) fulfill this requirement, and only for the *Translator* module. Interoperability is currently resolved by explicit technical translations from one service model to another. By this way, interoperability is only resolved at a technology level. On a more theoretical and formal level, the use of semantic and ontology based languages[1] is not sufficient to make service composition fully interoperable. Very often, service providers use different ontology domain and ontology transformations from one domain to another are more than needed.

Spontaneous management is only considered by MySIM[17] middleware. Indeed all of the other five middleware are goal-oriented and respond mainly to predefined functionality requests coming from the application layer. None of these middleware propose a spontaneous service composition that deliver new services and functionalities into the environment, without the intervention of users or applications. MySIM[17] proposes a service integration middleware that generates new services in the environment spontaneously and automatically. Compatible services are composed on the fly, without any intervention and upon the middleware own decision based on semantic and syntactic service matching.

The middleware listed above, do not propose solutions to address the problem of security or trust. They rely on the existing mechanisms proposed by the middleware and network layers, if any. Several other studies[14][15] address security features for service composition using contracts[15], verification formal methods[14], or a security model for enforcing access control in extensible component platforms[20].

No real autonomous composition management is provided. The middleware do not propose mechanism to manage their resources, functions, security, and performance, in face of failures and changes, with little or no human intervention. Pervasive environments that are capable of composing functionalities autonomously are still at preliminary state of consumption. A major domain that dealt with autonomous management of the composition is the multi-agent systems. Combining multi-agent systems and service-oriented architecture is a well known research field to add autonomy features to services[9][18][21][23].

## 5. Conclusions

The development of pervasive computing has put the accent on a well identified problem, the service composition problem. Composing services together on various platforms, extending environments with new functionalities, are the new trends pervasive computing aims to achieve. Many composition middleware have reached a certain maturity, and propose complete architectures and protocols to discover and compose services in pervasive environments. Many surveys[5][7][22][31][33] list service composition middleware according to predefined criteria or properties. They very often consider middleware for the composition of a particular technology such as Web services composition middleware. The application of service composition middleware to pervasive environment is rather new, and a real lack in analyzing and classifying service composition middleware under a reference model is noticed.

In this article, we surveyed six complete service composition architectures for pervasive environments, located in the middleware layer, MySIM[17], PERSE[30], SeSCo[10], Broker[6], SeGSeC[8] and WebDG[12]. We do not claim the exhaustiveness of our classification, but we think that the major middleware for service composition in pervasive environments are depicted. We introduced a novel approach to study the service composition problem. We studied these systems by reducing the composition problem to four main problems: the service translations, the composition plan generations, the plan contextual evaluations, and finally the real composition implementation. In each of these domains, several trends appeared to be commonly used: simple translation between diverse service technologies for the *Translator*, graph based approach or language composition one for the *Generator*, formal methods approach for the *Evaluator*, and discovery and invocation mechanisms for the *Builder*. Finally, we classified these middleware under several requirements related to the ubiquity of the environments. If some requirements such as discoverability and context awareness are well verified, others are still being explored such as interoperability, adaptability and QoS management. Security, spontaneous and autonomous management open the way to many promising research trends, at the intersection of several major domains such as artificial intelligence and autonomic computing, for service composition middleware in pervasive environments.

## References

[1] T. Bittner and M. Donnelly and S. Winter:, **Ontology and semantic interoperability**, CRCpress (Tailor & Francis), D. Prosperi and S. Zlatanova (ed.): Large-scale 3D data integration: Challenges and Opportunities, pages 139-160, 2005.

[2] T. Erl:, **Service-Oriented Architecture (SOA): Concepts, Technology, and Design**, Prentice Hall, 2005.

[3] B. Cole-Gomolski:, **Messaging Middleware Initiative Takes a Hit**, PComputerworld, 1997.

[4] Matjaz Juric and Poornachandra Sarang and Benny Mathew:, **Business Process Execution Language for Web Services (2nd edition)**, PACKT Publishing, 2006.

[5] A. Alamri and M. Eid and A. El Saddik "Classification of the state-of-the-art dynamic web services composition techniques", **Int. J. Web and Grid Services**, Vol. 2, No. 2, 2006, pp. 148-166.

[6] D. Chakraborty and A. Joshi and T. Finin and Y. Yesha "Service Composition for Mobile Environments", **Journal on Mobile Networking and Applications**, Special Issue on Mobile Services, Vol. 10, No. 4, 2005, pp. 435-451.

[7] S. Dustdar and W. Schreiner "A survey on web services composition", **Int. J. Web and Grid Services**, Vol. 1, No. 1, 2005, pp. 1-30.

[8] K. Fujii and T. Suda "Semantics-based dynamic service composition", **IEEE Journal on Selected Areas in Communications**, Vol. 23, No. 12, 2005.

[9] F. Ishikawa and N. Yoshioka and S. Honiden "Mobile agent system for Web service integration in pervasive network", **Systems and Computers in Japan**, Wiley-Interscience, Vol. 36, No. 11, 2005, pp. 34-48.

[10] S. Kalasapur and M. Kumar and B. Shirazi "Dynamic Service Composition in Pervasive Computing", **IEEE Transactions on Parallel and Distributed Systems**, Vol. 18, No. 7, 2007, pp. 907-918.

[11] B. Medjahed and Y. Atif "Context-based matching for

web service composition", **Distributed and Parallel Databases, Special Issue on Context-Aware Web Services**, Vol. 21, No. 1, 2006, pp. 5-37.

[12] B. Medjahed and A. Bouguettaya and A. K. Elmagarmid "Composing Web services on the Semantic Web", **The VLDB Journal**, Vol. 12, No. 4, 2003, pp. 333-351.

[13] M. Satyanarayanan "Pervasive Computing: Vision and Challenges", **IEEE Personal Communication**, 2001.

[14] G. Barthe and D. Gurov and M. Huisman "Compositional Verification of Secure Applet Interactions", in **FASE '02: Proceedings of the 5th International Conference on Fundamental Approaches to Software Engineering**, 2002, London, UK, pp. 15-32.

[15] N. Dragoni and F. Massacci and C. Schaefer and T. Walter and E. Vetillard "A Security-by-contracts Architecture for Pervasive Services", in **Security Privacy and Trust in Pervasive and Ubiquitous Computing Worshop**, 2007.

[16] B. Girma and L. Brunie and J.-M. Pierson "Planning-Based Multimedia Adaptation Services Composition for Pervasive Computing", in **2nd International Conference on Signal-Image Technology & Internet & based Systems (SITIS'2006)**, 2006, LNCS series, Springer Verlag.

[17] N. Ibrahim and F. Le Mouël and S. Frénot "MySIM: a Spontaneous Service Integration Middleware for Pervasive Environments", in **ACM International Conference on Pervasive Services (ICPS'2009)**, 2009, London, UK.

[18] Z. Maamar and S. Kouadri and H. Yahyaoui "A Web services composition approach based on software agents and context", in **SAC'04: Proceedings of the 2004 ACM symposium on Applied computing**, 2004, Nicosia, Cyprus.

[19] E. Niemela and J. Latvakoski "Survey of requirements and solutions for ubiquitous software", in **3rd international conference on Mobile and ubiquitous multimedia**, 2004, Vol. x, pp. 71-78.

[20] P. Parrend and S. Frénot "Component-Based Access Control: Secure Software Composition through Static Analysis", in **Proceedings of the 7th International Symposium**, 2008 Springer, LNCS 4954, Budapest, Hungary.

[21] C. Preist and C. Bartolini and A. Byde "Agent-based service composition through simultaneous negotiation in forward and reverse auctions", in **EC '03: Proceedings of the 4th ACM conference on Electronic commerce**, 2003.

[22] J. Rao and X. Su "A Survey of Automated Web Service Composition Methods", in **First International Workshop on Semantic Web Services and Web Process Composition**, 2004, SWSWPC, San Diego, California, USA.

[23] Q. B. Vo and L. Padgham "Conversation-Based Specification and Composition of Agent Services", in **Cooperative Information Agents (CIA)**, 2006, Edinburgh, UK, pp. 168-182.

[24] Z. Yang and R. Gay and C. Miao and J.-B. Zhang and Z. Shen and L. Zhuang and H. M. Lee "Automating integration of manufacturing systems and services: a semantic Web services approach", in **Industrial Electronics Society (IECON) 31st Annual Conference of**

IEEE, 2005.

[25] Ion Constantinescu and Boi Faltings and Walter Binder "Large Scale, Type-Compatible Service Composition", in **ICWS '04: Proceedings of the IEEE International Conference on Web Services**, Washington, DC, USA, 2004.

[26] Evren Sirin and James Hendler and Bijan Parsia "Semi-automatic Composition of Web Services using Semantic Descriptions", in **Web Services: Modeling, Architecture and Infrastructure Workshop**, Angers, France, 2003.

[27] Fabio Casati and Ski Ilnicki and Li-jie Jin and Vasudev Krishnamoorthy and Ming-Chien Shan "Adaptive and Dynamic Service Composition in eFlow", in **CAiSE '00: Proceedings of the 12th International Conference on Advanced Information Systems Engineering**, London, UK, 2000.

[28] Tao Gu and Hung Keng Pung and Da Qing Zhang "A Middleware for Building Context-Aware Mobile Services", in **Proceedings of IEEE Vehicular Technology Conference**, Los Angeles, USA, 2004.

[29] Shankar R. Ponnekanti and Armando Fox "SWORD: A developer toolkit for web service composition", in **11th World Wide Web Conference**, Honolulu, USA, 2002.

[30] S. Ben Mokhtar, "Semantic Middleware for Service-Oriented Pervasive Computing", Ph.D. thesis, University of Paris 6, Paris, France, 2007.

[31] D. Kuropka and H. Meyer "Survey on Service Composition", **Technical Report, Hasso-Plattner-Institute, University of Potsdam**, number 10, 2005.

[32] J. Floch ed. "Theory of adaptation", **Delivrable D2.2, Mobility and ADaptation enAbling Middleware (MADAM)**, 2006.

[33] C. Mascolo and S. Hailes and L. Lymberopoulos and P. Picco and P. Costa and G. Blair and P. Okanda and T. Sivaharan and W. Fritsche and M. and M. A. Rónai and K. Fodor and A. Boulis "Survey of Middleware for Networked Embedded Systems", **Technical Report, FP6 IP Project: Reconfigurable Ubiquitous Networked Embedded Systems**, 2005.

[34] T. Salminen, "Lightweight middleware architecture for mobile phones", Ph.D. thesis, Department of Electrical and Information Engineering, University of oulu, Oulu, Finland, 2005.

[35] UPnP Forum, "Understanding UPnP: A White Paper", **Technical Report,** 2000.

[36] The OWL Services Coalition, "OWL-S: Semantic Markup for Web Servicesr", **White paper,** 2003.

**Noha Ibrahim** holds an engineering diploma from Ecole Nationale Supérieure d'Informatique et de Mathématique Appliquée de Grenoble (ENSIMAG), and a Phd degree from National Institute for Applied Science (INSA) Lyon, France. The Phd is about service integration in pervasive environments. Her Phd focused on providing a spontaneous service integration middleware adapted for pervasive middleware. Her main interests are middleware for pervasive and ambient computing. Noha Ibrahim is currently a post doctoral at the Grenoble Informatics Laboratory where she works on service composition based framework for optimizing queries.

**IJCSI**

**Frédéric Le Mouël** holds an engineering diploma in Languages and Operating Systems, and a Phd degree from the University of Rennes 1, France. His dissertation focused on an adaptive environment for distributed execution of applications in a mobile computing context. Frédéric Le Mouël is currently associate professor in the National Institute for Applied Sciences of Lyon(INSA Lyon, France), Telecommunications Department, Center for Innovation in Telecommunication and Integration of Services (CITI Lab.). His main interests are service-oriented middleware and more specifically on the fields of dynamic adaptation, composition, coordination and trust of services.

**IJCSI**